\pdfoutput=1
\documentclass[aps,prd,preprint,superscriptaddress,tightenlines,nofootinbib]{revtex4}

\long\def\symbolfootnote[#1]#2{\begingroup%
\def\thefootnote{\fnsymbol{footnote}}\footnote[#1]{#2}\endgroup}

\usepackage{amsfonts,amsmath,amsthm,amssymb}
\usepackage{mathrsfs}
\usepackage{bm}
\usepackage{color}
\usepackage{epsfig}

\newcommand{\PRE}[1]{{#1}}   

\newcommand{\beq}{\begin{equation}}
\newcommand{\eeq}{\end{equation}}
\newcommand{\bea}{\begin{flushleft} \begin{eqnarray}}
\newcommand{\eea}{\end{eqnarray}\end{flushleft}}

\newcommand{\comment}[1]{}

\newcommand{\ci}[1]{}

\newcommand{\lb}{\left(}
\newcommand{\rb}{\right)}

\newcommand{\ba}{\begin{eqnarray}}
\newcommand{\ea}{\end{eqnarray}}
\newcommand{\be}{\begin{equation}}
\newcommand{\ee}{\end{equation}}
\newcommand{\bay}[1]{\left(\begin{array}{#1}}
\newcommand{\eay}{\end{array}\right)}

\newcommand{\ie}{\textit{i.e.}, }

\newcommand{\zt}[1]{\rm{#1}}

\def\xg{{\gamma}}
\def\xG{{\Gamma}}

\def\xvp{{\varphi}}
\def\xs{{\sigma}}

\usepackage[usenames,dvipsnames]{xcolor}
\definecolor{orange}{cmyk}{0,0.5,1,0}
\definecolor{rossoCP3}{cmyk}{0,.88,.77,.40}
\definecolor{graa}{rgb}{0.8,0.8,0.8}
\definecolor{blaa}{rgb}{0.2,0.2,0.6}

\begin{document}

\preprint{
\hfil
\begin{minipage}[t]{3in}
\begin{flushright}
\vspace*{.4in}
MPP--2015--315\\
LMU-ASC 81/15\\
\end{flushright}
\end{minipage}
}

\title{\PRE{\vspace*{0.9in}} \color{rossoCP3}{750~GeV diphotons from
    closed string states 
}}

\author{{\bf Luis A. Anchordoqui}}

\affiliation{Department of Physics and Astronomy,\\  Lehman College, City University of
  New York, NY 10468, USA
\PRE{\vspace*{.05in}}
}

\affiliation{Department of Physics,\\
 Graduate Center, City University
  of New York,  NY 10016, USA
\PRE{\vspace*{.05in}}
}

\affiliation{Department of Astrophysics,\\
 American Museum of Natural History, NY
 10024, USA
\PRE{\vspace*{.05in}}
}

\affiliation{Departamento de F\'{\i}sica,\\ Universidad Nacional de La
  Plata, C.C. 67, (1900) La Plata, Argentina
\PRE{\vspace*{.05in}}
}

\author{{\bf Ignatios Antoniadis}}
\affiliation{LPTHE, UMR CNRS 7589\\
Sorbonne Universit\'es, UPMC Paris 6, 75005 Paris, France
\PRE{\vspace*{.05in}}}

\affiliation{Albert Einstein Center, Institute for Theoretical Physics\\
Bern University, Sidlerstrasse 5, CH-3012 Bern, Switzerland
\PRE{\vspace*{.05in}}}

\author{{\bf Haim \nolinebreak Goldberg}}
\affiliation{Department of Physics,\\
Northeastern University, Boston, MA 02115, USA
\PRE{\vspace*{.05in}}
}

\author{{\bf Xing Huang}}
\affiliation{Department of Physics, \\
National Taiwan Normal University, Taipei, 116, Taiwan
\PRE{\vspace*{.05in}}
}

\author{{\bf Dieter L\"ust}}

\affiliation{Max--Planck--Institut f\"ur Physik, \\ 
 Werner--Heisenberg--Institut,
80805 M\"unchen, Germany
\PRE{\vspace*{.05in}}
}

\affiliation{Arnold Sommerfeld Center for Theoretical Physics 
Ludwig-Maximilians-Universit\"at M\"unchen,
80333 M\"unchen, Germany
\PRE{\vspace{.05in}}
}

\author{{\bf Tomasz R. Taylor}}

\affiliation{Department of Physics,\\
 Northeastern University, Boston, MA 02115, USA 
 \PRE{\vspace*{.05in}}
}

\begin{abstract}
  \noindent We show that low-mass-scale string compactifications, with
  a generic D-brane configuration that realizes the standard
  model by open strings, can explain the relatively broad peak in the
  diphoton invariant mass spectrum at 750~GeV recently reported by the
  ATLAS and CMS collaborations.  Under reasonable assumptions, we
  demonstrate that the excess could originate from a closed string
  (possibly axionic) excitation $\varphi$  that has a coupling
  with gauge kinetic terms. We estimate the $\varphi$ production rate
  from photon-photon fusion in elastic $pp$ scattering, using the
  effective photon and narrow width approximations. For string scales
  above todays lower limit  $M_s \approx 7~{\rm TeV}$, we can
  accommodate the diphoton rate observed at Run II while maintaining
  consistency with Run I data. 
\end{abstract}

\maketitle

Very recently, ATLAS~\cite{ATLAS} and CMS~\cite{CMS:2015dxe} announced
preliminary results on inclusive diphoton searches using
(respectively) $3.2~{\rm fb}^{-1}$ and $2.6~{\rm fb}^{-1}$ of data
recorded at a center-of-mass energy $\sqrt{s} = 13~{\rm TeV}$. The two
experiments observed an excess of events over expectations from
standard model (SM) processes in the invariant mass spectrum at
$\approx 750~{\rm GeV}$. This could be interpreted as decays of new
massive particle $\varphi$. For a narrow width approximation
hypothesis, ATLAS gives a local significance of $3.6\sigma$ and a
global significance of $2.0\sigma$ when accounting for the
look-elsewhere-effect in the mass range $M_\varphi/{\rm GeV} \in [200
- 2000]$. Signal-plus-background fits were also implemented assuming a
large width for the signal component. The most significant deviation
from the background-only hypothesis is reported for a mass of about
750~{\rm GeV} and a width $\Gamma_{\rm total} \approx 45~{\rm
  GeV}$. The local and global significances evaluated for the large
width fit are roughly 0.3 higher than that for the narrow width
approximation fit, corresponding to $3.9\sigma$ and $2.3\sigma$,
respectively. CMS reports a local significance of $2.6\sigma$ and a
global significance smaller than $1.2\sigma$. The observed number of
events corresponds to a production rate of $\sigma (pp \to \varphi +
{\rm anything}) \times {\cal B} (\varphi \to \gamma \gamma) \approx 3
- 6~{\rm fb}$.  The data at $\sqrt{s} = 13~{\rm TeV}$ prefer a cross
section which is roughly 16 times larger than the one at $\sqrt{s} =
8~{\rm TeV}$~\cite{Aad:2015mna,Khachatryan:2015qba}.  A wide-eyed fit
to the $pp \to \gamma \gamma$ rates demonstrates that the data at
$\sqrt{s} = 8~{\rm TeV}$ are incompatible with those at $\sqrt{s} =
13~{\rm TeV}$ at 95\% CL if the cross section grows less than about a
factor of 3.5~\cite{Franceschini:2015kwy}. Note that the background
from SM processes, which is dominated by $q \bar q \to \gamma \gamma$,
increases by a smaller factor: $\sigma (pp \to \gamma \gamma)
\approx 6~{\rm fb}$ at $\sqrt{s} = 8~{\rm TeV}$ and $\sigma (pp \to \gamma \gamma)
\approx 14~{\rm fb}$ at $\sqrt{s} = 13~{\rm TeV}$, after impossing
$M_{\gamma\gamma} > 750~{\rm GeV}$ and standard cuts.  Even though the excess is
not statistical significant yet, it is interesting to entertain the
possibility that it corresponds to a real signal of new physics. 

A plethora of models have been proposed to explain the
data including some string inspired
scenarios~\cite{Heckman:2015kqk,Cvetic:2015vit}. 
Actually in~\cite{Heckman:2015kqk}  the  new
massive particle $\varphi$ corresponds to a state on a `hidden' sector brane in F-theory that couples to the SM gauge bosons via (open string) messengers between the visible and hidden sector branes; on the other hand, 
in~\cite{Cvetic:2015vit}  $\varphi$ corresponds to an exotic open string state
on the visible sector branes. One feature of most of this kind of open string explanations is that  the
necessary  coupling $\varphi F^2$ is a priori forbidden by the $U(1)$
charge conservation in the open string D-brane sector, unless the
$U(1)$ is spontaneously broken by the expectation value of $\varphi$
that has Yukawa couplings to the messengers. ($F$ is the photon field strength.)
  
Herein we put forward an alternative solution from string theory,
namely that $\varphi$ corresponds to a closed string state.  Namely we consider extensions of the SM
based on D-brane string compactifications with large extra
dimensions~\cite{Antoniadis:1998ig}. The basic unit of gauge
invariance for D-brane constructions is a $U(1)$ field, so that a
stack of $N$ identical D-branes eventually generates a $U(N)$ theory
with the associated $U(N)$ gauge group; for $N = 2$, the gauge group
can be $Sp(1) \cong SU(2)$ rather than
$U(2)$~\cite{Blumenhagen:2005mu,Blumenhagen:2006ci}. In the presence
of many D-brane types, the gauge group becomes a product form $\prod
U(N_i)$, where $N_i$ reflects the number of D-branes in each stack. In
the perturbative regime, gauge interactions emerge as excitations of
open strings ending on D-branes, with gauge bosons due to strings
attached to stacks of D-branes and chiral matter due to strings
stretching between intersecting D-branes.

Our explanation of the peak in the diphoton invariant mass spectrum is
independent of the structure of the D-brane model. Nevertheless, to
motivate the discussion we adopt a minimal model containing 4 stacks
of D-branes. The basic setting of the gauge theory is given by $U(3)_C
\times Sp(1)_L \times U(1)_L \times
U(1)_{I_R}$~\cite{Cremades:2003qj,Anchordoqui:2011eg,Anchordoqui:2015uea}. In
the bosonic sector the open strings terminating on the QCD stack
contain, in addition to the $SU(3)_C$ octet of gluons $g_\mu^a$, an
extra $U(1)$ boson $C_\mu$, most simply the manifestation of a gauged
baryon number. The $Sp(1)_L$ stack is a terminus for the electroweak
gauge bosons $W^a_\mu$. The $U(1)_Y$ boson $Y_\mu$ that gauges the
usual electroweak hypercharge symmetry is a linear combination of
$C_\mu$, and the $U(1)$ bosons $B_\mu$ and $X_\mu$ terminating on the
separate $U(1)_L$ and $U(1)_{I_R}$ branes. The general properties of
the chiral spectrum are summarized in Table~\ref{table}.

\begin{table}
  \caption{Chiral  spectrum of SM fields in the 4 stack D-brane
    model. We have added the right handed neutrino stretching between
    the lepton brane and the right brane.}
\begin{center}
\begin{tabular}{ccccccc}
\hline
\hline
~~~Fields~~~ & ~~~Sector~~~  & ~~~Representation~~~ & ~~~$Q_B$~~~ & ~~~$Q_L$~~~ & ~~~$Q_{I_R}$~~~ & ~~~$Q_Y$~~~ \\
\hline
 $U_R$ &   $\phantom{^*}3 \leftrightharpoons 1^*$ &  $(3,1)$ & $1$ & $\phantom{-}0$ & $\phantom{-} 1$ & $\phantom{-}\frac{2}{3}$  \\[1mm]
  $D_R$ & $3 \leftrightharpoons 1$  & $( 3,1)$&    $1$ & $\phantom{-}0$ & $- 1$ & $-\frac{1}{3}$   \\[1mm]
  $L_L$ & $4 \leftrightharpoons 2$ &  $(1,2)$&    $0$ &  $\phantom{-}1$ & $\phantom{-}0$ & $-\frac{1}{2}$ \\[1mm]
  $E_R$ & $4 \leftrightharpoons 1$ &  $(1,1)$&   $0$ & $\phantom{-}1$ &  $- 1$ & $- 1$ \\[1mm]
 $Q_L$ & $3 \leftrightharpoons 2$ &  $(3,2)$& $1$ & $\phantom{-}0 $ & $\phantom{-} 0$ & $\phantom{-} \frac{1}{6}$    \\[1mm]
   $N_R$  &  $\phantom{^*}4 \leftrightharpoons 1^*$  &   $(1,1)$& $0$ & $\phantom{-}1$ & 
$\phantom{-} 1$ & $\phantom{-} 0$ \\ [1mm]
$H$ & $2 \leftrightharpoons 1$ &  $(1,2)$ & $0$ & $\phantom{-}0$ & $\phantom{-}1 $ &
$\phantom{-}\frac{1}{2}$  \\ [1mm]
\hline
\hline
\label{table}
\end{tabular}
\end{center}
\end{table}

One can check by inspection that the hypercharge,
\begin{equation}
Q_Y = \frac{1}{2} Q_{I_R} + \frac{1}{6} Q_B - \frac{1}{2} Q_L  \,, 
\label{hyperchargeY}
\end{equation}
is anomaly free. However, the $Q_B$ (gauged baryon number) is not
anomaly free and we expect this anomaly to be canceled via a
Green-Schwarz mechanism involving the exchange of twisted
Ramond-Ramond (RR) closed string
states~\cite{Green:1984sg,Witten:1984dg,Dine:1987bq,Lerche:1987qk,Ibanez:1999it}.
There is an explicit mass term in the Lagrangian for the new gauge
field $-\frac{1}{2} {M'}^2 Y'_\mu {Y'}^\mu$ whose scale comes from the
compactification scheme. The scalar that gets eaten up to give the
longitudinal polarization of the $Y'$ is a closed string field and
there is no extra Higgs particle~\cite{Ghilencea:2002da}.  In addition
to the intermediate RR field, which is absorbed by the $Y'$ in the
anomaly cancellation, there is a closed string mode $\varphi$ which
couples to the anomaly free combination of the hypercharge (\ref{hyperchargeY}).
It can be either a scalar field from the Neveu-Schwarz sector that is
complexified with the RR state absorbed by $Y'$, or another RR
pseudo-scalar (axion) coupled to $F\tilde{F}$.

In this Letter we propose that the observed diphoton excess originates
from the closed string excitation $\xvp$.  There are two properties of
the scalar $\varphi$ that are necessary for explaining the 750~GeV
signal. It should be a special closed string state with dilaton-like
or axion-like coupling to $F^2$ (respectively to $F \tilde F$) of the
electromagnetic field, but {\em decoupled} from $F^2$ of color
$SU(3)$.  The couplings of closed string states to gauge fields do
indeed distinguish between different D-brane stacks, depending on the
localization properties of D-branes with respect to $\varphi$ in the
compact dimensions.  More specifically, it is quite natural to assume
that $\varphi$ is a closed string mode that is associated to the
wrapped cycles of the $U(1)_L$ and $U(1)_{I_R}$ stack of D-branes,
however is not or only weakly attached to the wrapped cycle of
$Sp(1)_L$ or the color $SU(3)$ stack of D-branes. In this way, we can
avoid unwanted dijet signals~\cite{comment}. Furthermore, since the string mass scale
is now known to be larger than $M_s \approx 7~{\rm
  TeV}$~\cite{Khachatryan:2015dcf}, the mass $M_{\varphi}\approx
750$~GeV must be suppressed with respect to the string scale by some
anomalous loop corrections. Because $\varphi$ is a twisted closed
string localized at an orbifold singularity, its coupling to $\gamma
\gamma$ should be suppressed by $M_s^{-1}$, provided the bulk is
large~\cite{Antoniadis:2002cs}. With this in mind, we parametrize the
coupling of $\xvp$ to the photon by the following vertex \be \frac
{c_{\xg\xg}} v \xvp F^2\,, \ee where $v \sim M_s$.  To remain in the
perturbative range, we also require $c_{\xg\xg}$ to be reasonably
small. The partial decay width of $\xvp$ to diphotons then follows as
\begin{equation}
\Gamma_{\gamma\gamma} = \frac{c_{\gamma\gamma}^2}{4\pi}
\frac{M_\xvp^3}{v^2}\, . \label{eq:photonwidth}
\end{equation}

\begin{figure}[t]
\begin{center}
\includegraphics[width=0.517\linewidth]{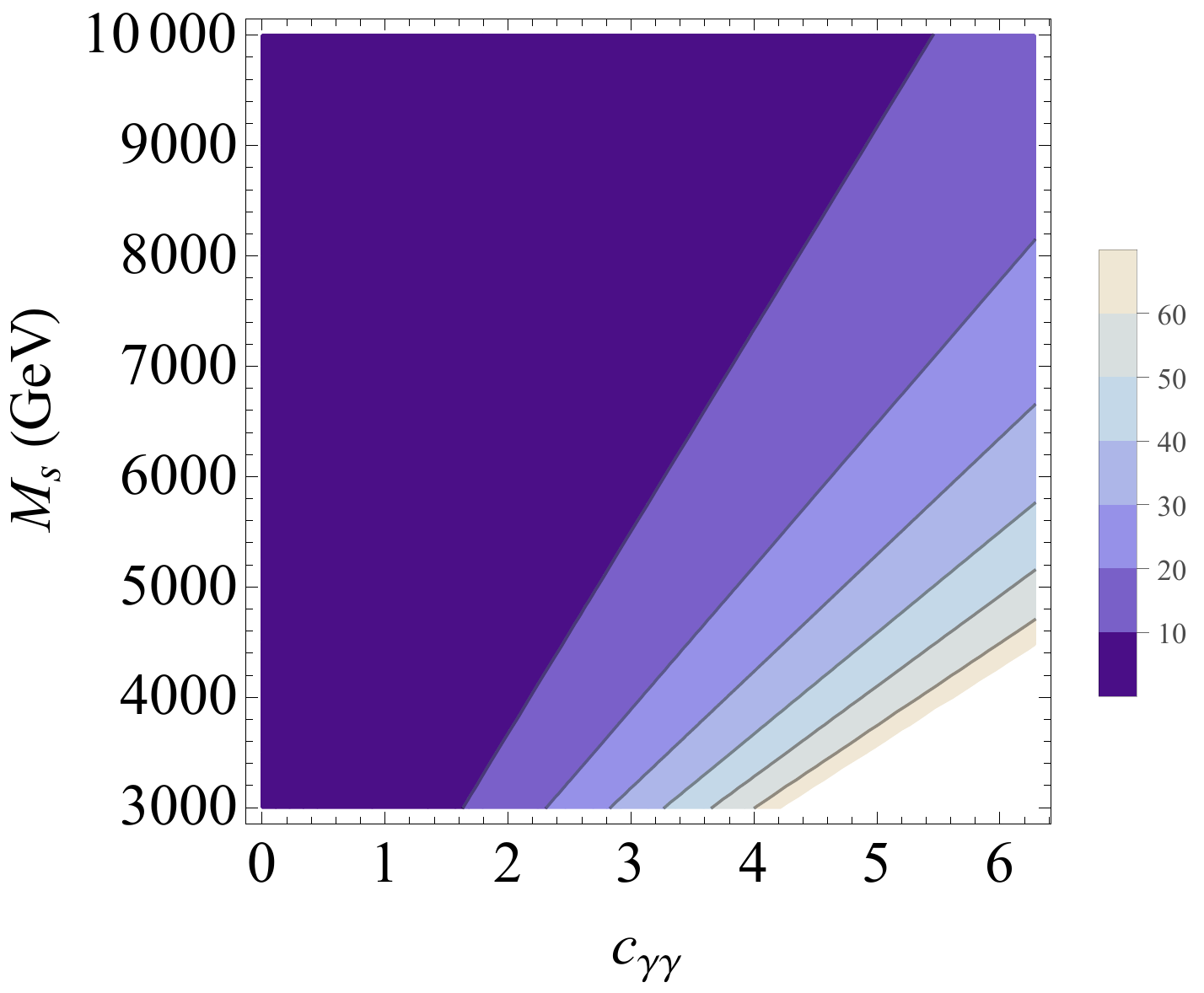}
\includegraphics[width=0.45\linewidth]{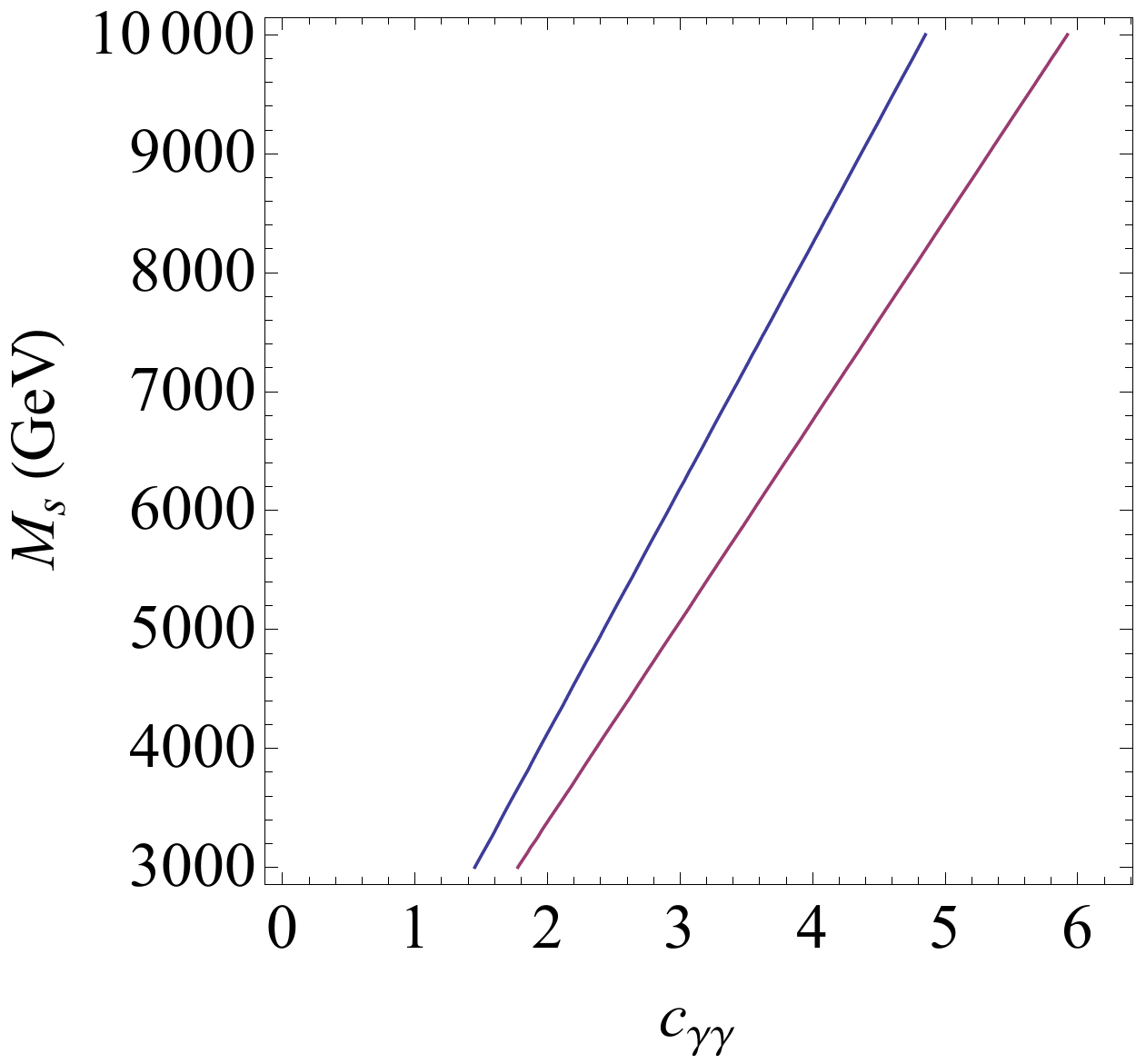}
\end{center}
\vspace{-0.3cm}
\caption[]{Contours of constant partial width $\xG_{\xg\xg}$ (left) and best fit contours ($\xs = 5$ fb) of cross section $pp \to pp \xvp \to pp \xg\xg$ (right), for $M_{\xvp} \simeq 750~{\rm GeV}$, $\Gamma_{\zt{total}}=45$ GeV and $\sqrt{s} = 13~{\rm TeV}$. The color encoded scales are in GeV. The blue and red contours (on the right) correspond to $q_* = 170$~MeV and $q_* = 130$~MeV, respectively.}
\label{fig1}
\end{figure}

The diphoton signal is produced via photon-photon fusion with $\xvp$ as the
resonance state~\cite{Fichet:2015vvy,Csaki:2015vek}. The simplest
way to get a reliable estimate of $\sigma (pp \to \gamma \gamma)$ is
provided by the equivalent photon approximation (originally due to Fermi~\cite{Fermi:1924tc} and later on developed
by Weizs\"acker~\cite{vonWeizsacker:1934nji} and
Williams~\cite{Williams:1934ad}). Under the narrow width approximation,
the cross section is found to be
\begin{equation}
\sigma(pp \to pp \gamma\gamma) = \frac{8 \pi^2}{M_\xvp}
\frac{\Gamma_{\gamma\gamma}^2}{\Gamma_{\zt{total}}} \int dx_1 \ dx_2 \
f^\gamma_s (x_1) \ f^\gamma_s (x_2) \ \delta( x_1 x_2 s - M_\xvp^2 )\,,
\end{equation}
where $f^\gamma_s (x_1)$ is the photon distribution function, which
for small $x$ takes the following approximate form \be f_s^\gamma(x)
dx = \frac{dx}{x} \frac{2 \alpha}{\pi} \log \left[ \frac{q_*}{m_p}
  \frac{1}{x} \right]\,, \ee where $\alpha \approx 1/129$, $m_p$ is
the proton mass, and $q_*$ is the inverse of the minimum impact
parameter for elastic
scattering~\cite{Budnev:1974de,Martin:2014nqa}. Following~\cite{Csaki:2015vek}
we consider the range $130~{\rm MeV} < q_* < 170~{\rm MeV}$, which 
accommodates the LHC two photon Higgs production cross section.
The total cross sections are \be \sigma_{\sqrt s = 13~\text{TeV}} =
162~\text{fb}~\left(\frac{\Gamma_{\zt{total}}}{45~\text{GeV}} \right)
{\cal B}^2(\xvp \rightarrow \gamma\gamma)\,, \ee for $q_* = 170$~MeV
and \be \sigma_{\sqrt s = 13~\text{TeV}} =
73~\text{fb}~\left(\frac{\Gamma_{\zt{total}}}{45~{\rm GeV}} \right)
{\cal B}^2(\xvp \rightarrow \gamma\gamma)\,, \ee for $q_* = 130~{\rm
  MeV}$. With the observed total decay width of $\xG_{\zt{total}} =
45~{\rm GeV}$, the branching fraction is given by \be {\cal B}^2(\xvp
\rightarrow \gamma\gamma) = \frac{2.3 \times 10^6 c_{\gamma
    \gamma}^2}{\pi v_{{\rm GeV}}^2}\, ,  \ee where
$v_{\rm GeV} \equiv v/{\rm GeV}$. We perform a scan in the parameter
space $(c_{\xg\xg}, v)$. As one can see in Fig.~\ref{fig1}, for
reasonably large $c_{\xg\xg}$, a string scale $M_s \sim v$ above the
experimental lower bound of 7~TeV~\cite{Khachatryan:2015dcf} can give
$\sigma_{\sqrt s = 13~\text{TeV}} \sim 5$~fb, \ie ${\cal B}(\xvp
\rightarrow \gamma\gamma) \sim 0.17$ ($q_* = 170~{\rm MeV}$) or ${\cal
  B}(\xvp \rightarrow \gamma\gamma) \sim 0.26$ ($q_* = 130~{\rm
  MeV}$). These correspond to $\Gamma_{\gamma \gamma} = 8~{\rm GeV}$
and $\Gamma_{\gamma \gamma} = 12~{\rm GeV}$, respectively. For a cross
section $\sigma_{\sqrt s = 13~\text{TeV}} \sim 6$~fb, $9~{\rm GeV}
\alt \Gamma_{\gamma \gamma} \alt 13~{\rm GeV}$.

The assumed coupling of $\varphi$ to the hypercharge field strength yields
additional decay channels in the visible sector, namely $\varphi \to \gamma
Z$ and $\varphi \to ZZ$, with
\begin{equation}
\frac{\Gamma_{\gamma Z}}{\Gamma_{\gamma \gamma}} = 2 \tan^2 \theta_W
\approx 0.6 \quad {\rm and} \quad \frac{\Gamma_{Z Z}}{\Gamma_{\gamma
    \gamma}} =  \tan^4 \theta_W \approx 0.08 \,.
\end{equation}
The precise branching fraction into $\gamma Z$ and $ZZ$ is one of the
hallmark predictions of our model. In general one would expect a
coupling to hypercharge and a coupling to $SU(2)$ bosons with a
parameter controlling their relative strength. This would appear in
the emergent couplings to $ZZ$, $\gamma \gamma$, and $\gamma Z$. This would be
the generic situation even in other effective field theory models.

One may wonder whether the missing fraction of the decay width can be
explained through the coupling of $\varphi$ to gravitons. However, as
we show in the Appendix, the KK tower of gravitons gives a negligible
contribution to the total width. On the other hand, the missing
fraction of the decay width could arise from the coupling of $\varphi$
to other bulk fields, such as fermions, which has less number of
derivatives. These hidden fermions could make a contribution to the
dark matter content of the
universe~\cite{Dienes:2011ja,Dienes:2011sa}.

For collisions at $\sqrt{s} = 8~{\rm TeV}$, \be \sigma_{\sqrt s =
    8~\text{TeV}} =
  31~\text{fb}~\left(\frac{\Gamma_{\zt{total}}}{45~\text{GeV}} \right)
  {\cal B}^2(\xvp \rightarrow \gamma\gamma)\,, \ee when $q^* =
  170~{\rm MeV}$ and \be \sigma_{\sqrt s = 8~\text{TeV}} =
  6.5~\text{fb}~\left(\frac{\Gamma_{\zt{total}}}{45~\text{GeV}}
  \right) {\cal B}^2(\xvp \rightarrow \gamma\gamma)\,, \ee when $q^* =
  130~{\rm MeV}$~\cite{Csaki:2015vek}. Although both agree with LHC8
  data~\cite{Aad:2015mna,Khachatryan:2015qba}, we see that smaller values of $q^*$ correspond to a much
  larger increase in going from 8~TeV to 13~TeV.

In closing, we comment on other anomalies observed by the LHC
experiments.  As noted elsewhere~\cite{Anchordoqui:2015uea} the new
Abelian gauge bosons of the D-brane structure (which suffer a mixed
anomaly with the SM but are made self consistent by the Green-Schwarz
mechanism) can also accommodate the
diboson~\cite{Aad:2015owa,Khachatryan:2014hpa} and
dijet~\cite{Aad:2014aqa,Khachatryan:2015sja} excesses above SM
backgrounds observed in the invariant mass region of $\approx 1.8 -
2.0~{\rm TeV}$ by ATLAS and CMS in collisions at $\sqrt{s} = 8~{\rm
  TeV}$. Under  reasonable assumptions we have shown that $\sigma
(pp \to Z') \times {\cal B} (Z' \to JJ) = 123~{\rm fb}$ and $\sigma
(pp \to Z') \times {\cal B} (Z' \to W^+W^-/ZZ) = 7.7~{\rm fb}$, in
agreement with observations. The $Z'$ production cross section grows
by a factor $\approx 7$ when going from $\sqrt{s} = 8$ to
13~TeV~\cite{Pelaggi:2015knk}. It is straightforward to see that the
predicted production of dijet topologies is in good agreement with the
small excess observed by ATLAS and CMS around
1.8~TeV~\cite{Khachatryan:2015dcf,ATLAS:2015nsi}, and that the
diboson~\cite{ATLAS:13-1,ATLAS:13-2,CMS:2015nmz} and
dilepton~\cite{ATLAS:13-3,CMS:2015nhc} final states are partially
consistent with data.

\acknowledgments{We thank Tere Dova and Carlos Garcia Canal for
  some valuable discussion. L.A.A.  is supported by U.S. National Science
  Foundation (NSF) CAREER Award PHY1053663 and by the National
  Aeronautics and Space Administration (NASA) Grant No. NNX13AH52G; he
  thanks the Center for Cosmology and Particle Physics at New York
  University for its hospitality.  H.G. and
  T.R.T. are supported by NSF Grant No. PHY-1314774.  X.H.  is
  supported by the MOST Grant 103-2811-M-003-024. D.L. is partially
  supported by the ERC Advanced Grant Strings and Gravity
  (Grant.No. 32004) and by the DFG cluster of excellence ``Origin and
  Structure of the Universe.''  Any opinions, findings, and
  conclusions or recommendations expressed in this material are those
  of the authors and do not necessarily reflect the views of the
  National Science Foundation.}

\section*{Appendix}

The $\xvp R^2$ vertex is at least suppressed by $M_s/M_{\zt{Pl}}$
compared to $\xvp F^2$, where $R$ is the Ricci scalar and $M_{\rm Pl}
\sim 10^{19}~{\rm GeV}$ is the Planck mass.  In fact it should be
$(M_\xvp /M_{\zt{Pl}})^2$ because there are two gravitons and four
derivatives.  To estimate the graviton emission we sum over KK modes with momentum less than
$M_\xvp$ (so that the decay is kinematically allowed). The number of
modes is given by the volume of a 6-sphere of radius $(M_\xvp L)/(2
\pi)$, where $L$ is the scale of the extra dimension and is related to
$M_{\zt{Pl}}$ by
\[
L^n M_s^{n+2} \sim M_{\zt{Pl}}^2\,.
\]
For $M_s \sim 10$~TeV and $n=6$, this implies $L \sim 10~{\rm GeV}^{-1}$. The graviton decay width can then be estimated as 
\[
\frac {\xG_{GG}} {\xG_{\xg\xg}} < \lb \frac {M_s} {M_{\zt{Pl}}} \rb^2 \frac {S_5} n \lb \frac {M_\xvp L}{2 \pi} \rb^6 \sim 10^{-11}\,,
\]
where $S_5$ is the area of unit 5-sphere. As we can see, this is basically negligible.


\begin{thebibliography}{99}
\bibitem{comment} We may note in passing that if an excess of dijet
  events is observed in future LHC data, this can be easily
  accommodated by coupling $\varphi$ to $F^2$ of $SU(3)_C$ changing
  the localization properties of D-branes with respect to $\varphi$ in
  the internal space.
\end{thebibliography}
\end{document}